\documentstyle[12pt,dina4]{article}

\def\title #1 {
   \headsep=1.0in
   \baselineskip=30pt
			\begin{center}
   {\titlebold #1}
   \end{center}
			\vskip .75in }
\def\author #1 {
   \baselineskip=30pt
   \begin{center}
   {\timeslarge #1}
   \end{center}
			\vskip .25in }
\def\address #1 {
   \baselineskip=24pt
   \begin{center}
   {\timesitalic #1}
   \end{center}
   \vskip 1.0in }

\def\disp {\displaystyle}

\def\conj #1 {\overline #1}

\def\be {\begin{equation}}
\def\ee {\end{equation}}

\def\ba {\begin{array}}
\def\ea {\end{array}}
\def\bea {\begin{eqnarray}}
\def\eea {\end{eqnarray}}

\def\et {$$}

\def\etn {$$}

\def\ett {$$}

\def\ettn{$$}

\def\eqn#1 {(\ref{#1}) }

\newdimen\twoeqncolwidth
\newdimen\twoeqncolwidtha
\newdimen\twoeqncolwidthb
\newdimen\twoeqncolsep
\newdimen\twoeqnlinset
\def\twoeqn#1&#2\et{
   \hbox to\twoeqnlinset{\hfil}
   \hbox to\twoeqncolwidth{$\disp#1$\hfil}
   \hbox to\twoeqncolsep{\hfil}
   \hbox to\twoeqncolwidth{$\disp#2$\hfil}\eqno{\rm (\theequation)}$$}
\def\twoeqnt#1&#2\ett{
   \hbox to\twoeqnlinset{\hfil}
   \hbox to\twoeqncolwidtha{$\disp#1$\hfil}
   \hbox to\twoeqncolsep{\hfil}
   \hbox to\twoeqncolwidthb{$\disp#2$\hfil}\eqno{\rm (\theequation)}$$}
\def\twoeqnn#1&#2\etn{
   \hbox to\twoeqnlinset{\hfil}
   \hbox to\twoeqncolwidth{$\disp#1$\hfil}
   \hbox to\twoeqncolsep{\hfil}
   \hbox to\twoeqncolwidth{$\disp#2$\hfil}\eqno\phantom{\rm
(\theequation)}$$}
\def\twoeqntn#1&#2\ettn{
   \hbox to\twoeqnlinset{\hfil}
   \hbox to\twoeqncolwidtha{$\disp#1$\hfil}
   \hbox to\twoeqncolsep{\hfil}
   \hbox to\twoeqncolwidthb{$\disp#2$\hfil}\eqno\phantom{\rm
(\theequation)}$$}

\def\rawpicture #1 by #2 (#3){
  \vbox to #2{
    \hrule width #1 height 0pt depth 0pt
    \vfill
    \special{picture #3} 
    }
  }
\def\scaledpicture #1 by #2 (#3 scaled #4){{
  \dimen0=#1 \dimen1=#2
  \divide\dimen0 by 1000 \multiply\dimen0 by #4
  \divide\dimen1 by 1000 \multiply\dimen1 by #4
  \rawpicture \dimen0 by \dimen1 (#3 scaled #4)}
  }

\def\beginparmode{\endmode
  \begingroup \def\endmode{\par\endgroup}}
\let\endmode=\par
{\obeylines\gdef\
{}}

\def\body			
  {\beginparmode}		

\def\head#1{			
  \goodbreak\vskip 0.5truein	
  {\immediate\write16{#1}
   \uppercase{#1}\par}
   \nobreak\vskip 0.25truein\nobreak}

\def\itemj{\par\hang\textindent}
\def\beginitems{
\par\medskip\bgroup\def\i##1 {\itemj{##1}}\def\ii##1 {\itemitem{##1}}
\leftskip=36pt\parskip=0pt}
\def\enditems{\par\egroup}

\def\beneathrel#1\under#2{\mathrel{\mathop{#2}\limits_{#1}}}

\def\refto#1{[#1]}		

\def\references			
  {\head{REFERENCES}		
   \beginparmode
   \frenchspacing \parindent=0pt \leftskip=1truecm
   \parskip=8pt plus 3pt \everypar{\hangindent=\parindent}}

\gdef\refis#1{\itemj{#1.\ }}			

\gdef\journal#1, #2, #3, 1#4#5#6{		
    {\sl #1~}{\bf #2} (1#4#5#6), #3 }		

\def\annp{\journal Ann. Phys. (N.Y.), }

\def\Annp{\journal Ann. Physik, }

\def\aspm{\journal Advanced Studies in Pure Mathematics, }

\def\cmp{\journal Comm. Math. Phys., }

\def\eurolett{\journal Europhysics Lett., }

\def\ijmpa{\journal Int. J. Mod. Phys. A, }

\def\ijmpb{\journal Int. J. Mod. Phys. B, }

\def\jappp{\journal J. Appl. Phys., }

\def\jphc{\journal J. Physique C, }

\def\jpI{\journal J. Physique I, }

\def\jpcoll{\journal J. Physique Coll, }

\def\jcr{\journal J. Chem. Res., }

\def\jetp{\journal Sov. Phys. JETP, }

\def\jetpl{\journal JETP Lett., }

\def\jpj{\journal J. Phys. Soc. Japan, }

\def\jmp{\journal J. Math. Phys., }

\def\jpa{\journal J. Phys. A, }

\def\jpc{\journal J. Phys. C, }

\def\jpcon{\journal J. Phys.: Condens. Matter, }

\def\ptp{\journal Prog. Theor. Phys., }

\def\jetp{\journal Sov. Phys. JETP, }

\def\jsp{\journal J. Stat. Phys., }

\def\lmp{\journal Lett. Math. Phys., }

\def\lnp{\journal Lecture Notes in Physics, }

\def\mpla{\journal Mod. Phys. Lett. A, }

\def\npb{\journal Nucl. Phys. B, }

\def\physica{\journal Physica, }

\def\pla{\journal Phys. Lett. A, }

\def\plb{\journal Phys. Lett. B, }

\def\prep{\journal Physics Reports, }

\def\pra{\journal Phys. Rev. A, }

\def\prb{\journal Phys. Rev. B, }

\def\prl{\journal Phys. Rev. Lett., }

\def\prs{\journal Proc. Roy. Soc. (London) A, }

\def\pr{\journal Phys. Rev., }

\def\rmp{\journal Rev. Mod. Phys., }

\def\sjnp{\journal Sov. J. Nucl. Phys., }

\def\tmp{\journal Theor. Math. Phys., }

\def\zpb{\journal Z. Phys. B, }

\def\zp{\journal Z. Phys., }

\def\reff#1{Ref.~#1}			
\def\Reff#1{Ref.~#1}			
\def\[#1]{[\refcite{#1}]}
\def\refcite#1{{#1}}
\def\(#1){(\call{#1})}
\def\call#1{{#1}}
\def\taghead#1{}
\def\frac#1#2{{#1 \over #2}}

\def\sla{\raise.15ex\hbox{$/$}\kern-.57em}
\def\leaderfill{\leaders\hbox to 1em{\hss.\hss}\hfill}
\def\twiddle{\lower.9ex\rlap{$\kern-.1em\scriptstyle\sim$}}
\def\bigtwiddle{\lower1.ex\rlap{$\sim$}}
\def\gtwid{\mathrel{\raise.3ex\hbox{$>$\kern-.75em\lower1ex\hbox{$\sim$}}}}
\def\ltwid{\mathrel{\raise.3ex\hbox{$<$\kern-.75em\lower1ex\hbox{$\sim$}}}}
\def\square{\kern1pt\vbox{\hrule height 1.2pt\hbox{\vrule width 1.2pt\hskip
3pt
   \vbox{\vskip 6pt}\hskip 3pt\vrule width 0.6pt}\hrule height
0.6pt}\kern1pt}
\def\tdot#1{\mathord{\mathop{#1}\limits^{\kern2pt\ldots}}}

\def\pmb#1{\setbox0=\hbox{#1}%
  \kern-.025em\copy0\kern-\wd0
  \kern  .05em\copy0\kern-\wd0
  \kern-.025em\raise.0433em\box0 }

\def\refto#1{[#1]}		

\def\references			
  {\section*{References}		
   \beginparmode
   \frenchspacing \parindent=0pt \leftskip=1truecm
   \parskip=8pt plus 3pt \everypar{\hangindent=\parindent}}

\def\endreferences{\body}

\def\reff#1{Ref.~#1}			
\def\Reff#1{Ref.~#1}			
\def\[#1]{[\refcite{#1}]}
\def\refcite#1{{#1}}
\def\refeq#1{(\ref{#1})}

\catcode`@=11
\newcount\r@fcount \r@fcount=0
\newcount\r@fcurr
\immediate\newwrite\reffile
\newif\ifr@ffile\r@ffilefalse
\def\w@rnwrite#1{\ifr@ffile\immediate\write\reffile{#1}\fi\message{#1}}

\def\writer@f#1>>{}
\def\referencefile{
  \r@ffiletrue\immediate\openout\reffile=\jobname.ref%
  \def\writer@f##1>>{\ifr@ffile\immediate\write\reffile%
    {\noexpand\refis{##1} = \csname r@fnum##1\endcsname = %
     \expandafter\expandafter\expandafter\strip@t\expandafter%
     \meaning\csname r@ftext\csname r@fnum##1\endcsname\endcsname}\fi}%
  \def\strip@t##1>>{}}

\def\citeall#1{\xdef#1##1{#1{\noexpand\refcite{##1}}}}
\def\refcite#1{\each@rg\citer@nge{#1}}

\def\each@rg#1#2{{\let\thecsname=#1\expandafter\first@rg#2,\end,}}
\def\first@rg#1,{\thecsname{#1}\apply@rg}	
\def\apply@rg#1,{\ifx\end#1\let\next=\relax
\else,\thecsname{#1}\let\next=\apply@rg\fi\next}

\def\citer@nge#1{\citedor@nge#1-\end-}
\def\citer@ngeat#1\end-{#1}
\def\citedor@nge#1-#2-{\ifx\end#2\r@featspace#1 
  \else\citel@@p{#1}{#2}\citer@ngeat\fi}	
\def\citel@@p#1#2{\ifnum#1>#2{\errmessage{Reference range #1-#2\space is
bad.}
    \errhelp{If you cite a series of references by the notation M-N, then M
and
    N must be integers, and N must be greater than or equal to M.}}\else%
 {\count0=#1\count1=#2\advance\count1
by1\relax\expandafter\r@fcite\the\count0,%
  \loop\advance\count0 by1\relax
    \ifnum\count0<\count1,\expandafter\r@fcite\the\count0,%
  \repeat}\fi}

\def\r@featspace#1#2 {\r@fcite#1#2,}	
\def\r@fcite#1,{\ifuncit@d{#1}
    \newr@f{#1}%
    \expandafter\gdef\csname r@ftext\number\r@fcount\endcsname%
                     {\message{Reference #1 to be supplied.}%
                      \writer@f#1>>#1 to be supplied.\par}%
 \fi%
 \csname r@fnum#1\endcsname}
\def\ifuncit@d#1{\expandafter\ifx\csname r@fnum#1\endcsname\relax}%
\def\newr@f#1{\global\advance\r@fcount by1%
    \expandafter\xdef\csname r@fnum#1\endcsname{\number\r@fcount}}

\let\r@fis=\refis			
\def\refis#1#2#3\par{\ifuncit@d{#1}
   \newr@f{#1}%
   \w@rnwrite{Reference #1=\number\r@fcount\space is not cited up to
now.}\fi%
  \expandafter\gdef\csname r@ftext\csname r@fnum#1\endcsname\endcsname%
  {\writer@f#1>>#2#3\par}}

\def\ignoreuncited{
   \def\refis##1##2##3\par{\ifuncit@d{##1}%
     \else\expandafter\gdef\csname r@ftext\csname
r@fnum##1\endcsname\endcsname%
     {\writer@f##1>>##2##3\par}\fi}}

\def\r@ferr{\endreferences\errmessage{I was expecting to see
\noexpand\endreferences before now;  I have inserted it here.}}
\let\r@ferences=\references
\def\references{\r@ferences\def\endmode{\r@ferr\par\endgroup}}

\let\endr@ferences=\endreferences
\def\endreferences{\r@fcurr=0
  {\loop\ifnum\r@fcurr<\r@fcount
    \advance\r@fcurr by
1\relax\expandafter\r@fis\expandafter{\number\r@fcurr}%
    \csname r@ftext\number\r@fcurr\endcsname%
  \repeat}\gdef\r@ferr{}\endr@ferences}

\let\r@fend=\endpaper\gdef\endpaper{\ifr@ffile
\immediate\write16{Cross References written on []\jobname.REF.}\fi\r@fend}

\catcode`@=12

\def\reftorange#1#2#3{[\refcite{#1}--\setbox0=\hbox{\refcite{#2}}\refcite{#3}]}

\citeall\refto		
\citeall\reff		%
\citeall\Reff		%

\ignoreuncited

\renewcommand{\title}[1]{\large\bf \mbox{}\\ \mbox{}\\ \mbox{}\\ \mbox{}\\
     #1\bigskip\medskip\\}
\renewcommand{\author}[1]{\large #1\\ \smallskip}
\renewcommand{\address}[1]{{\narrower\normalsize\it #1\\}\bigskip}

\def\phi{\varphi}
\def\-{{\bf --}}

\newcommand{\la}{\lambda}
\newcommand{\eps}{\varepsilon}

\newcounter{num}

\def\cre#1#2{c^+_{#1 #2}}
\def\ann#1#2{c_{#1 #2}}
\def\num#1#2{n_{#1 #2}}

\def\e{{\rm e}}

\def\qLam#1#2{q^{#1}\left({i\over 2}#2 \eta\right)}
\def\qLam2#1#2{q^{#1}\left({i\over 2}#2 {\eta\over 2}\right)}

\def\i{\hbox{i}}

\begin{document}
\begin{center}

\titlepage

\title{A new integrable two parameter model of
strongly correlated electrons in one dimension\footnote{Work
performed within the research program of the
Sonderforschungsbereich 341, K\"oln-Aachen-J\"ulich}}

\vskip1cm

\author{R. Z. Bariev\footnote{Permanent address: The Kazan Physico-Technical
Institute of the Russian Academy of Sciences, Kazan 420029, Russia},
A. Kl\"umper, J. Zittartz}

\address{Institut f\"ur Theoretische Physik,
Universit\"at zu K\"oln, Z\"ulpicher Str. 77,\\
D-50937 K\"oln 41, Germany.\footnote{Email:
kluemper@thp.uni-koeln.de, zitt@thp.uni-koeln.de}}

\end{center}
\vskip1cm

\begin{abstract}
A new one-dimensional fermion model depending on two independent interaction
parameters is formulated and solved exactly by the Bethe ansatz method.
The Hamiltonian of the model contains the Hubbard interaction and correlated
hopping as well as pair hopping terms. The density-density and pair
correlations are calculated which manifest superconducting properties in
certain regimes of the phase diagram.
\end{abstract}

PACS {71.28.+d,74.20.-z,75.10.Lp}

\vspace{1cm}

\today

\bibliographystyle{alpha}

\newpage

The study of strongly correlated fermion systems has attracted considerable
attention during the last decade. In particular the discovery of high
temperature superconductivity \refto{BednorzM86}
has led to the investigation of various
electronic mechanisms for an explanation of this phenomenon. In one
space dimension several new integrable models have been found exhibiting
different physical behaviour \reftorange{Schlott87}{BaresB90,
EssKS9293,BarievKSZ93sum,Bariev94b,BarievKSZ94a,BarievKSZmult}{Karn94}.
Notably the model with correlated hopping \refto{BarievKSZ93sum},
the anisotropic $tJ$ model \refto{BarievKSZ94a}, and models with pair
hopping processes \refto{BarievKSZmult} show
dominant superconducting correlations.

Due to integrability many properties
of these models can be found nonperturbatively by an exact solution on the
basis of a Bethe ansatz \reftorange{Yang67}{LiebWu68,Suth75}{Suth}
and conformal field theory \refto{BelaPZ84,Card86a}.
On the other hand integrability imposes severe conditions on the
interaction parameters of a model. Usually, integrable cases are restricted
to discrete points or depend on one parameter only. Therefore, the study of
competition of different interaction terms is difficult.

In this letter we report on a new integrable two parameter model of strongly
correlated electrons unifying several previously known integrable systems
such as the correlated hopping model and the $tJ$ model. The phase diagram
is very rich with a crossover from a regime with dominating density-density
correlations to a regime with dominating (superconducting) pair correlations.

Our starting point is a one-dimensional Hamiltonian with correlated
hopping terms, with the Hubbard on site interaction, and with
pair hopping processes
\bea
H=&-&\sum_{j,\sigma}\left(\cre{j}{\sigma}\ann{j+1}{\sigma}+
\cre{j+1}{\sigma}\ann{j}{\sigma}\right)
\exp\left[-{1\over 2}(\eta-\sigma\gamma)\num{j}{,-\sigma}
-{1\over 2}(\eta+\sigma\gamma)\num{j+1}{,-\sigma}\right]\cr
&+&\sum_{j}\left[U\num{j}{\uparrow}\num{j}{\downarrow}+
t_p\left(\cre{j}{\uparrow}\cre{j}{\downarrow}\ann{j+1}{\downarrow}\ann{j+1}{\uparrow}+h.c.\right)\right],
\label{hamil}
\eea
where $j$ denotes the sites and we use
standard notation for fermion operators. Tight binding Hamiltonians of
this type have been considered since a long time
\refto{Hubbard,Hirsch89a} and have
been studied as effective one-band Hamiltonians for the
description of cuprate superconductors, see \refto{SimAl93,BoerKS95} and
references therein.
In \refeq{hamil} we have
included an anisotropy $\gamma$ in the correlated hopping term and
also a pair hopping term $t_p$. Obviously, the integrable Hubbard chain
\refto{LiebWu68} is a special
case of \refeq{hamil} with $\eta=\gamma=t_p=0$. For $\eta=\pm\gamma$ and
$t_p=U=0$
we recover the correlated hopping model
\refto{BarievKSZ93sum}. The exact solution for the particular
case $\eta=\ln 2$, $\gamma=0$ was studied in \refto{Karn94,Karn95}.

Here we shall establish the integrability of \refeq{hamil} under the
condition
\be
t_p=U/2=\eps\left[2\e^{-\eta}(\cosh\eta-\cosh\gamma)\right]^{1/2},\qquad\eps=\pm 1,
\label{cond}
\ee
leaving us with two continuous parameters $\eta$, $\gamma$
($|\eta|>\gamma\ge 0$ for hermiticity), and the
discrete sign $\eps$ where $\eps=\pm 1$ corresponds to repulsive
and attractive pair processes, respectively.
The exact solution for the eigenstates and eigenvalues
of Hamiltonian \refeq{hamil} can be obtained from a Bethe ansatz.
The structure of the Bethe ansatz equations follows from the solution of the
two particle scattering problem.
The nonvanishing elements of the $S$ matrix are
\bea
S_{11}^{11}(\la)&=&S_{22}^{22}(\la)=1,\cr
S_{12}^{12}(\la)&=&S_{21}^{21}(\la)={\sin \la\over\sin(\la-i\gamma)},\cr
S_{21}^{12}(\la)&=&{i\sinh\gamma\over\sin(\la-i\gamma)}\e^{-i\la},\cr
S_{12}^{21}(\la)&=&{i\sinh\gamma\over\sin(\la-i\gamma)}\e^{i\la},
\label{Smat}
\eea
where $\la=\la_1-\la_2$, and the $\la_j$ are suitable particle
rapidities related to the momenta of the electrons by
\be
k(\la)=\cases{
\Theta(\la,a),& $\eps=+1$,\cr
\pi-\Theta(\la,a),& $\eps=-1$,\cr}
\label{moment}
\ee
with the function $\Theta$ and the parameter $a$ defined by
\bea
\Theta(\la,a)&=&2\arctan\left(\coth a \tan{\la}\right),\cr
a&=&{1\over 4}\left\{\ln\left[{\sinh{1\over 2}(\eta+\gamma)
\over\sinh{1\over 2}(\eta-\gamma)}\right]-\gamma\right\}.\label{parm}
\eea
A necessary and sufficient condition for the applicability of the
Bethe ansatz are the Yang-Baxter equations \refto{Yang67,Baxt82b}.
These equations are satisfied
by \refeq{Smat}, i.e. by the $S$ matrix of \refeq{hamil} under the condition
\refeq{cond}.

The Bethe ansatz equations are derived following the standard procedure by
imposing periodic boundary conditions. Each state of the Hamiltonian is
specified by a set of particle rapidities $\la_j$, $j=1,$..., $N$, and
a set of spin rapidities $\Lambda_\alpha$, $\alpha=1,$..., $N_\downarrow$,
where
$N$ is the total number of electrons and $N_\downarrow$ the number of down spin
electrons. All rapidities within a given set have to be different,
corresponding to Fermi statistics. They have to satisfy the Bethe ansatz
equations
\bea
\left[{\sin\left(\la_j-ia\right)\over
\sin\left(\la_j+ia\right)}\right]^L&=&
\prod_{\alpha=1}^{N_\downarrow}
{\sin\left(\la_j-\Lambda_\alpha+i\gamma/2\right)\over
\sin\left(\la_j-\Lambda_\alpha-i\gamma/2\right)},\cr
\prod_{j=1}^{N}
{\sin\left(\Lambda_\alpha-\la_j+i\gamma/2\right)\over
\sin\left(\Lambda_\alpha-\la_j-i\gamma/2\right)}
&=&-\prod_{\beta=1}^{N_\downarrow}
{\sin\left(\Lambda_\alpha-\Lambda_\beta+i\gamma\right)\over
\sin\left(\Lambda_\alpha-\Lambda_\beta-i\gamma\right)},
\label{BA}
\eea
where $L$ is the number of lattice sites.
The total energy and momentum of the system are given in terms of the
particle rapidities $\la_j$ as
\bea
E&=&-2\sum_{j=1}^N\cos k_j\cr
&=&2 \eps\sum_{j=1}^N
\left[\cosh 2a-{\sinh^22a\over\cosh 2a-\cos 2\la_j}\right],\cr
P&=&\sum_{j=1}^Nk(\la_j).
\label{energ}
\eea

Next we discuss the physical properties of the model. First consider
the limit $\eta\to -\infty$, $\eps=+1$.
In this limit double occupations on the same site are excluded
and the remaining dynamics is identical to that of the supersymmetric
$tJ$ model \refto{Schlott87,BaresB90,BarievKSZ94a} as can be seen
from the Bethe ansatz equations and also from a canonical transformation
of the Hamiltonian
\be
H_{\rm eff}=\e^AH\e^{-A},\label{eff}
\ee
generated by $A=\e^{-|\eta|/2}(d-d^+)$ where
\be
d=\sum_{l<m,\sigma=\pm 1}(-1)^{l-m}P_{l+1,m} \cre{l+1}{\sigma}\ann{l}{\sigma}
(1-\num{l}{,-\sigma}).
\ee
$P_{a,b}$ denotes the cyclic shift operator acting on the sites
$a$, $a+1$, ..., $b$, moving the state at site $a$ to site $b$.
In first order of $A$ the effective Hamiltonian \refeq{eff} does not
involve interactions describing transitions between states with
doubly occupied sites and those without doubly occupied sites.
In the limit $\eta\to -\infty$ the resulting Hamiltonian is that of
the supersymmetric $tJ$ Hamiltonian. A more detailed account
will be given in a separate publication.

Secondly, a particle-hole
transformation $T$ together with a sublattice rotation
($\cre{j}{\sigma}\to(-1)^j\ann{j}{\sigma}$) applied to $H(\eta)$
(\ref{hamil},\ref{cond})
yields a resulting Hamiltonian of the same form, namely
\be
T H(\eta) T^{-1}=\e^\eta H(-\eta)+(L-N)U.\label{trafo}
\ee
Therefore we can restrict the further investigation to one specific sign of
$\eta$. We choose $\eta>0$ corresponding to repulsive (physical) correlated
hopping of the electrons and positive values for $a$ \refeq{parm}.

For the groundstate the magnetization is zero, $N_\downarrow=N/2$, and
the second set of equations in \refeq{BA} can be solved for the
$\Lambda_\alpha$ in terms of the $\la_j$ similar to the treatment in
\refto{BarievKSZ93sum} (Appendix of the second paper).
Inserting the result into
the first set of equations in \refeq{BA} and introducing the density function
$\rho(\la)$ for the distribution of $\la_j$ in the thermodynamic limit,
we obtain the linear integral equation
\be
2\pi\rho(\la)={2\sinh 2a\over\cosh 2a-\cos 2\la}+
\int_I\varphi(\la-\mu)\rho(\mu)d\mu,\label{inteq}
\ee
where
\be
\phi(\la)=1+2\sum_{n=1}^\infty\e^{-n\gamma}{\cos 2n\la\over\cosh n\gamma}.
\label{phi}
\ee
In order to minimize the groundstate energy
\be
{E_0\over L}=2 \eps\int_I
\left[\cosh 2a-{\sinh^22a\over\cosh 2a-\cos 2\la}\right]\rho(\lambda)
d\la,\label{erg}
\ee
the integration interval $I$ in (\ref{inteq},\ref{erg}) has to be chosen
symmetrically
around 0 ($I=[-K,+K]$) or $\pi/2$ ($I=[\pi/2-K,\pi/2+K]$), for $\eps=+1$
or $-1$, respectively. The parameter $K$ is determined by the subsidiary
condition for the total density $\rho=N/L$ of the electrons
\be
\int_I\rho(\la)d\la=\rho.\label{den}
\ee

Results for the groundstate energy per lattice site $E_0$ obtained by
numerical solutions of these equations are presented in Figs. 1 and 2 for
various values of $\eta$ and $\gamma$. For large values of the
correlated hopping parameter $\eta$ the processes described by Hamiltonian
\refeq{hamil} simplify drastically. The only single particle process allowed
is the hopping of electrons from singly occupied sites to empty sites with
hopping rate $t=-1$. In addition, pair hopping processes occur from doubly
occupied sites to empty sites with parameters $t_p=U/2=\eps$. For $\eps=+1$
the groundstate consists of only single electrons for densities $0<\rho<1$,
with dependence of $E_0$ on $\rho$ as for free spinless fermions
($E_0=-{2\over\pi}\sin\pi\rho$). Upon further filling doubly occupied
sites are created. However, the system freezes in as all hopping processes
are suppressed. All states with arbitrary distributions of singly and
doubly occupied sites are eigenstates of the Hamiltonian where the energy
is just given by the Hubbard interaction.
Therefore, $E_0$ depends linearly on $\rho$ for $1<\rho<2$,
see Fig. 1a). For the case $\eps=-1$ the formation of doubly occupied sites is
favoured for all densities. The dynamics of the pairs is described by free
spinless fermions ($E_0=-{2\over\pi}\sin\pi{\rho\over 2}-\rho$)
for all densities $\rho$, see Fig. 2a).

We next study the long distance behaviour of correlation functions by means of
finite size studies and application of conformal field theory. For details of
this method the reader is referred to \refto{BelaPZ84,Card86a,BarievKSZ93sum}.
There are two different kinds of excitations in model
(\ref{hamil},\ref{cond}). One
type of excitations corresponds to gapless particle-hole excitations, i.e. a
redistribution of the $\la_j$ parameters, the second type consists of spin
excitations
with gap. The latter excitations become gapless for $\gamma=0$ which situation
will be studied in a separate publication. The central charge of model
(\ref{hamil},\ref{cond}) is $c=1$. The long-distance behaviour of the
density-density and pair correlations is given by
\be
\langle\rho(r)\rho(0)\rangle\simeq\rho^2+A_1r^{-2}+A_2r^{-\alpha}\cos(2k_Fr),
\ee
and
\be
\langle
\cre{r}{\uparrow}\cre{r}{\downarrow}\ann{0}{\downarrow}\ann{0}{\uparrow}\rangle\simeq
B r^{-\beta},
\ee
where
\be
2k_F=\pi\rho.
\ee
The exponents $\alpha$ and $\beta$ describing the algebraic decay are
calculated from the dressed charge function $\xi(\la)$ which has to satisfy
the integral equation
\be
2\pi\xi(\la)=2\pi+
\int_I\varphi(\la-\mu)\xi(\mu)d\mu.\label{xi}
\ee
We obtain
\be
\alpha=1/\beta=\xi(K)^2/2
\ee
for $\eps=+1$, and a similar equation with $K$ replaced by $\pi/2+K$ for
$\eps=-1$.

The one-particle Green's function
$\langle\cre{r}{\sigma}\ann{0}{\sigma}\rangle$ shows exponential decay
since hole excitations have a mass gap. Consequently the momentum
distribution function $\langle\num{k}{\sigma}\rangle$ is analytic in
$k$, in particular there is no singularity at $k_F$.

The model does not show finite off-diagonal long-range order. However,
we observe a longer range of the pair correlations in comparison to
density-density correlations for certain regimes of interactions and
particle density. Numerical results for the dependence of the exponent
$\beta$ on the density $\rho$ are shown in Figs. 1 and 2. For all values
of the interaction parameters $\eta$ and $\gamma$ there is a crossover from
a regime with dominant density-density correlations ($1<\beta<2$) to a regime
with dominant pair correlations ($\beta<1$) at a ``critical density" $\rho_c$.
For repulsive pair processes ($\eps=+1$) the size of
regime ``$\beta<1$" is shrinking
for increasing values of $\eta$ and decreasing values of $\gamma$
($\rho_c\to 2$ for $\gamma\to 0$), see Figs. 1a) and b).
For attractive pair processes ($\eps=-1$) regime ``$\beta<1$" is growing
for increasing values of $\eta$ and decreasing values of $\gamma$
($\rho_c\to 0$ for $\gamma\to 0$ and $\eta\to\infty$), see Figs. 2a) and b).

In summary we have presented a new integrable model for strongly correlated
electron systems depending on two free parameters.
We have obtained exact results for the groundstate energy
of the system as well as the critical exponents for correlation functions
showing generically a crossover between regimes with
dominant density-density correlations and pair correlations, respectively.

After completion of this work we learnt about the formal proof
of integrability of model \refeq{hamil} in the particular case $\gamma=0$
\refto{Brack94,Beduerf95}.

\section*{Acknowledgments}
One of the authors (RZB) gratefully acknowledges the hospitality of the
Institut f\"ur Theoretische Physik der Universit\"at zu K\"oln as well as
partial financial support by the International Science Foundation grant no.
NNU000.

\newpage
\def\and{and\ }
\def\eds{eds.\ }
\def\edi{ed.\ }

\references

\def\mtb{M. T. Batchelor}
\def\rjb{R. J. Baxter}
\def\dk{D. Kim}
\def\pap{P. A. Pearce}
\def\nyr{N. Yu. Reshetikhin}
\def\ak{A. Kl\"umper}

\refis{AbramS64} M. Abramowitz, I. A. Stegun, ``Handbook of
Mathematical Functions", Washington, U.S. National Bureau of Standards 1964;
New York, Dover 1965.

\refis{Affl86} I. Affleck, \prl 56, 746, 1986

\refis{AKLT87} I. Affleck, T. Kennedy, E. H. Lieb \and H. Tasaki, \prl 59,
799, 1987

\refis{AKLT88} I. Affleck, T. Kennedy, E. H. Lieb \and H. Tasaki, \cmp 115,
477, 1988

\refis{AfflGSZ89} I. Affleck, D. Gepner, H. J. Schulz \and T. Ziman,
\jpa 22, 511, 1989

\refis{AfflM88} I. Affleck \and J.B. Marston, \jpc 21, 2511, 1988

\refis{AkutDW89} Y. Akutsu, T. Deguchi \and M. Wadati, in Braid Group, Knot
Theory and Statistical
Mechanics, \eds C. N. Yang \and M. L. Ge, World Scientific, Singapore, 1989

\refis{AkutKW86a} Y. Akutsu, A. Kuniba \and M. Wadati,\jpj 55, 1466, 1986

\refis{AkutKW86b} Y. Akutsu, A. Kuniba \and M. Wadati,\jpj 55, 2907, 1986

\refis{AlcaBB87} F. C. Alcaraz, M. N. Barber \and \mtb,\prl 58, 771, 1987

\refis{AlcaBB88} F. C. Alcaraz, M. N. Barber \and \mtb,\annp 182, 280, 1988

\refis{AlcaBGR88} F. C. Alcaraz, M. Baake, U. Grimm \and V. Rittenberg,
\jpa 21, L117, 1988

\refis{AlcaM89}  F. C. Alcaraz \and M. J. Martins, \jpa 22, 1829, 1989

\refis{AlcaM90}  F. C. Alcaraz \and M. J. Martins, \jpa 23, 1439-51, 1990

\refis{Alex75} S. Alexander, \pla 54, 353-4, 1975

\refis{AndrBF84} G. E. Andrews, \rjb\ \and P. J. Forrester, \jsp 35, 193,
1984

\refis{And87} P. W. Anderson, Science 235, 1196, 1987

\refis{And90} P. W. Anderson, \prl 64, 1839, 1990

\refis{ArrAl94} L. Arrachea \and A. A. Aligia, \prl 73, 2240, 1994

\refis{BDV82} O. Babelon, H. J. de Vega, \and C. M. Viallet, \npb 200 [FS4],
266, 1982

\refis{BarbBP87} \mtb, M.N. Barber \and \pap,\jsp 49, 1117, 1987

\refis{BarbB89} M.N. Barber \and \mtb, \prb 40, 4621, 1989

\refis{Barb91} M.N. Barber, \physica A 170, 221, 1991

\refis{BaresB90} P. A. Bares \and G. Blatter, \prl 64, 2567, 1990

\refis{Bariev8182} R. Z. Bariev, \tmp 49, 261, 1981; 1021, 1982

\refis{Bariev82} R. Z. Bariev, \tmp 49, 1021, 1982

\refis{Bariev91} R. Z. Bariev, \jpa 24, L549, 1991; L919, 1991

\refis{BarievKSZ93} R. Z. Bariev, A. Kl\"{u}mper, A. Schadschneider
\and J. Zittartz, \jpa 26, 1249, 1993; 4863

\refis{BarievKSZ93sum} R. Z. Bariev, A. Kl\"{u}mper, A. Schadschneider
\and J. Zittartz, \jpa 26, 1249, 1993; 4863; \physica B 194-196, 1417, 1994

\refis{BarievKSZ94b} R. Z. Bariev, A. Kl\"{u}mper, A. Schadschneider
\and J. Zittartz, \prb 50, 9676, 1994

\refis{BarievKSZ94a} R. Z. Bariev, A. Kl\"{u}mper, A. Schadschneider
\and J. Zittartz, \zpb 96, 395, 1995

\refis{BarievKSZmult} R. Z. Bariev, A. Kl\"{u}mper, A. Schadschneider
\and J. Zittartz, \prb 50, 9676, 1994; \jpa, in press, 1995

\refis{ZittKSB94} J. Zittartz, A. Kl\"{u}mper, A. Schadschneider
\and R. Z. Bariev, \physica B 194-196, 1417, 1994

\refis{Bariev94a} R. Z. Bariev, \prb 49, 1474, 1994

\refis{Bariev94b} R. Z. Bariev, \jpa 27, 3381, 1994

\refis{BarouchM71} E. Barouch \and B. M. McCoy, \pra 3, 786, 1971

\refis{Baxt70} \rjb,\jmp 11, 3116, 1970

\refis{Baxt71b} \rjb,\prl 26, 834, 1971

\refis{Baxt72} \rjb,\annp 70, 193, 1972

\refis{Baxt73} \rjb,\jsp 8, 25, 1973

\refis{Baxt80} \rjb,\jpa 13, L61--70, 1980

\refis{Baxt81a} \rjb,\physica 106A, 18--27, 1981

\refis{Baxt81b} \rjb,\jsp 26, 427--52, 1981

\refis{Baxt82a} \rjb,\jsp 28, 1, 1982

\refis{Baxt82b} \rjb, ``Exactly Solved Models in Statistical Mechanics",
Academic Press, London, 1982.

\refis{BaxtP82} \rjb\space \and \pap,\jpa 15, 897, 1982

\refis{BaxtP83} \rjb\space \and \pap,\jpa 16, 2239, 1983

\refis{BazhR89} V.V. Bazhanov \and \nyr,\ijmpa 4, 115--42, 1989

\refis{BazhB93} V.V. Bazhanov \and \rjb,\physica A 194, 390--396, 1993

\refis{BednorzM86} J. G. Bednorz \and K. A. M\"uller, \zpb 64, 189, 1986

\refis{Beduerf95} G. Bed\"urftig \and H. Frahm, preprint, ITP-UH-02/95

\refis{BelaPZ84} A. A. Belavin, A. M. Polyakov \and A. B. Zamolodchikov,
\npb 241, 333, 1984

\refis{Bethe31} H. A. Bethe,\zp 71, 205, 1931

\refis{BlotCN86} H. W. J. Bl\"ote, J. L. Cardy \and M. P. Nightingale, \prl
56, 742,
1986

\refis{BogK89} N. M. Bogoliubov \and V. E. Korepin, \ijmpb 3, 427-439, 1989

\refis{BoerKS95} J. de Boer, V. E. Korepin, A. Schadschneider, \prl 74,
789, 1995

\refis{BogIR86} N. M. Bogoliubov, A.\ G.\ Izergin \and
N.\ Y.\ Reshetikhin, \jetpl 44, 405, 1986

\refis{Brack94} A. J. Bracken, M. D. Gould, J. R. Links \and Y.-Z. Zhang,
preprint, cond-mat/9410026

\refis{BretzD71} M. Bretz \and J. G. Dash, \prl 27, 647, 1971

\refis{Bretz77} M. Bretz, \prl 38, 501, 1977

\refis{Buy86} W. J. L. Buyers, R. M. Morra, R. L. Armstrong, P. Gerlach
\and K. Hirakawa, \prl 56, 371, 1986

\refis{EssKS9293} F. H. L. Essler, V. E. Korepin \and K.Schoutens,
\prl 68, 2960, 1992; \prl 70, 73, 1993

\refis{KawUO89} N. Kawakami, T. Usuki \and A. Okiji, \pla 137, 287, 1989

\refis{KawY90} N. Kawakami \and S.-K. Yang, \prl 65, 2309, 1990

\refis{KawY91} N. Kawakami \and S.-K. Yang, \prb 44, 7844, 1991

\refis{Kaw93} N. Kawakami, \prb 47, 2928, 1993

\refis{KorBI93} V. E. Korepin, N.M. Bogoliubov, \and A.G. Izergin,
``Quantum Inverse Scattering Method and Correlation
Functions", Cambridge University Press, 1993.

\refis{Morra88} R. M. Morra, W. J. L. Buyers, R. L. Armstrong \and K. Hirakawa,
\prb 38, 543, 1988

\refis{SimAl93} M. E. Simon \and A. A. Aligia, \prb 48, 7471, 1993

\refis{ShasS90} B. S. Shastry \and B. Sutherland, \prl 66, 243, 1990

\refis{Shastry88} B. S. Shastry, \jsp 50, 57, 1988

\refis{Stei87} M. Steiner, K. Kakurai, J. K. Kjems, D. Petitgrand \and R. Pynn,
\jappp 61, 3953, 1987

\refis{Tun90} Z. Tun, W. J. L. Buyers, R. L. Armstrong, K. Hirakawa \and
B. Briat, \prb 42, 4677, 1990

\refis{Tun91} Z. Tun, W. J. L. Buyers, A. Harrison \and J. A. Rayne, \prb 43,
13331, 1991

\refis{Ren87} J. P. Renard, M. Verdaguer, L. P. Regnault, W. A. C. Erkelens,
J. Rossa-Mignod \and W. G. Stirling, \eurolett 3, 945, 1987

\refis{Ren88} J. P. Renard, M. Verdaguer, L. P. Regnault, W. A. C. Erkelens,
J. Rossa-Mignod, J. Ribas, W. G. Stirling \and C. Vettier, \jappp 63, 3538,
1988

\refis{Reg89} L. P. Regnault, J. Rossa-Mignod, J. P. Renard, M. Verdaguer
\and C. Vettier, \physica B 156 \& 157, 247, 1989

\refis{Colom87} P. Colombet, S. Lee, G. Ouvrard \and R. Brec, \jcr, 134, 1987

\refis{deGroot82} H. J. M. de Groot, L. J . de Jongh, R. D. Willet \and
J. Reeyk, \jappp 53, 8038, 1982

\refis{Capp88} A. Cappelli, Recent Results in Two-Dimensional Conformal
Field
Theory, in Proceedings of the XXIV International Conference on High Energy
Physics,
\eds R. Kotthaus \and J. K\"uhn, Springer, Berlin, 1988

\refis{CappIZ87a} A. Cappelli, C. Itzykson \and J.-B. Zuber, \npb {280
[FS18]},
445--65, 1987

\refis{CappIZ87b} A. Cappelli, C. Itzykson \and J.-B. Zuber, \cmp 113,
1--26, 1987

\refis{Card84a} J. L. Cardy, \jpa 17, L385, 1984

\refis{Card86a} J. L. Cardy, \npb {270 [FS16]}, 186, 1986

\refis{Card86b} J. L. Cardy, \npb {275 [FS17]}, 200, 1986

\refis{Card88} J. L. Cardy, ``Phase Transitions and Critical Phenomena,
Vol.11",
\eds C. Domb \and J.L. Lebowitz, Academic Press, London 1988

\refis{Card89} J. L. Cardy, Conformal Invariance and Statistical Mechanics,
in Les
Houches, Session XLIV, Fields, Strings and Critical Phenomena, \eds E.
Br\'ezin \and
J. Zinn-Justin, 1989

\refis{ChoiKK90} J.-Y. Choi, K. Kwon \and D. Kim, \eurolett xx, to appear,
1990

\refis{ChoiKP89} J.-Y. Choi, D. Kim \and \pap, \jpa 22, 1661--71, 1989

\refis{CvetDS80} D. M. Cvetkovic, M. Doob \and H. Sachs, ``Spectra of
Graphs", Academic Press, London 1980

\refis{DateJKMO87} E. Date, M. Jimbo, A. Kuniba, T. Miwa, \and M. Okado,
\npb
B290, 231--273, 1987

\refis{DateJKMO88} E. Date, M. Jimbo, A. Kuniba, T. Miwa, \and M. Okado,
\aspm 16,
17, 1988

\refis{DateJMO86} E. Date, M. Jimbo, T. Miwa \and M. Okado,\lmp 12, 209,
1986

\refis{DateJMO87} E. Date, M. Jimbo, T. Miwa \and M. Okado,\prb 35, 2105--7,
1987

\refis{DaviP90} B. Davies \and \pap, \ijmpb {}, this issue, 1990

\refis{deVeK87} H. J. de Vega \and M. Karowski, \npb {285 [FS19]}, 619, 1987

\refis{deVeW85} H. J. de Vega \and F. Woynarovich,\npb 251, 439, 1985

\refis{deVeW90} H. J. de Vega \and F. Woynarovich,\jpa 23, 1613, 1990

\refis{DestdeVeW92} C. Destri \and H. J. de Vega,\prl 69, 2313, 1992

\refis{diFrSZ87} P. di Francesco, H. Saleur \and J.-B. Zuber, \jsp 49,
57--79, 1987

\refis{diFrZ89} P. di Francesco \and J.-B. Zuber, $SU(N)$ Lattice Models
Associated
with Graphs, Saclay preprint SPhT/89-92, 1989

\refis{DijkVV88} R. Dijkgraaf, E. Verlinde \and H. Verlinde, in Proceedings
of the
1987 Copenhagen Conference, World Scientific, 1988

\refis{DijkVVV89} R. Dijkgraaf, C. Vafa, E. Verlinde \and H. Verlinde,\cmp
123, 485, 1989

\refis{DombG76} ``Phase Transitions and Critical Phenomena, Vol.6",
Academic Press, London 1976

\refis{FateZ85} V. A. Fateev \and A. B. Zamolodchikov, \jetp 62, 215, 1985

\refis{FendG89} P. Fendley \and P. Ginsparg, \npb 324, 549--80, 1989

\refis{FodaN89} O. Foda \and B. Nienhuis, \npb {},{},1989

\refis{FoersK93} A. Foerster \and M. Karowski, \npb 408 [FS], 512, 1993

\refis{FrieQS84} D. Friedan, Z. Qiu \and S. Shenker, \prl 52, 1575, 1984; in
``Vertex Operators in Mathematics and Physics", \eds J. Lepowsky, S.
Mandelstam \and
I.M. Singer, Springer, 1984

\refis{FrahmK90} H. Frahm \and V. E. Korepin, \prb 42, 10553, 1990

\refis{FrahmYF90} H. Frahm, N.-C. Yu \and M. Fowler, \npb 336, 396, 1990

\refis{Frei93} W.-D. Freitag, Dissertation, Universit\"at zu K\"oln, 1993

\refis{Gaudin83} M. Gaudin, ``La Fonction d'Onde de Bethe", Masson, Paris,
1983.

\refis{GepnQ87} D. Gepner \and Z. Qiu, \npb 285, 423--53, 1987

\refis{Gins88} P. Ginsparg,\npb {295 [FS21]}, 153--70, 1988

\refis{Gins89a} P. Ginsparg, Applied Conformal Field Theory, in Les
Houches,
Session XLIV, Fields, Strings and Critical Phenomena, \eds E. Br\'ezin \and
J.
Zinn-Justin, 1989

\refis{Gins89b} P. Ginsparg, Some Statistical Mechanical Models and
Conformal Field
Theories, Trieste Spring School Lectures, HUTP-89/A027

\refis{GradR80} I.S. Gradshteyn \and I.M. Ryzhik, ``Tables of Integrals,
Series and Products", Academic
Press, New York, 1980.

\refis{Grif72} R. B. Griffiths, ``Phase Transitions and Critical Phenomena,
Vol.1",\eds C. Domb \and M. S. Green, Academic Press, London 1972

\refis{Hald83a} F. D. M. Haldane, \prl 50, 1153, 1983

\refis{Hald83b} F. D. M. Haldane, \pla 93, 464, 1983

\refis{Hame85} C. J. Hamer,\jpa 18, L1133, 1985

\refis{Hame86} C. J. Hamer,\jpa 19, 3335, 1986

\refis{Hirsch89a} J. E. Hirsch, \pla 134, 451, 1989

\refis{Hirsch89b} J. E. Hirsch, \physica  C 158, 326, 1989

\refis{Hubbard} J. Hubbard, \prs 276, 238, 1963

\refis{HuiD93} A. Hui \and S. Doniach, \pr 48, 2063, 1993

\refis{Huse82} D. A. Huse,\prl 49, 1121--4, 1982

\refis{Huse84} D. A. Huse, \prb 30, 3908, 1984

\refis{Idz94} M. Idzumi, T. Tokihiro \and M. Arai, \jpI 4, 1151, 1994

\refis{ItzySZ88} C. Itzykson, H. Saleur \and J-B. Zuber, ``Conformal
Invariance and Applications to
Statistical Mechanics", World Scientific, Singapore, 1988

\refis{JimbM84} M. Jimbo \and T. Miwa, \aspm 4, 97--119, 1984

\refis{JimbMO87} M. Jimbo, T. Miwa \and M. Okado, \lmp 14, 123--31, 1987

\refis{JimbMO88} M. Jimbo, T. Miwa \and M. Okado, \cmp 116, 507--25, 1988

\refis{JimbMT89} M. Jimbo, T. Miwa \and A. Tsuchiya,``Integrable Systems in
Quantum Field Theory and
Statistical Mechanics", \aspm 19, ,1989

\refis{JohnKM73} J.D. Johnson, S. Krinsky, \and B.M. McCoy,\pra 8, 2526,
1973

\refis{Kac79} V. G. Kac, \lnp 94, 441--445, 1979

\refis{KadaB79} L. P. Kadanoff \and A. C. Brown, \annp 121, 318--42, 1979

\refis{Karo88} M. Karowski, \npb {300 [FS22]}, 473, 1988

\refis{Karn94} I. N. Karnaukhov, \prl 73, 1130, 1994

\refis{Karn95} I. N. Karnaukhov, \prb 51, ?, 1995

\refis{Kato87} A. Kato, \mpla 2, 585, 1987

\refis{KimP87} \dk\space \and \pap,\jpa 20, L451--6, 1987

\refis{KimP89}  \dk\space \and \pap,\jpa 22, 1439--50, 1989

\refis{Kiri89} E. B. Kiritsis, \plb  217, 427, 1989

\refis{KiriR86} A. N. Kirillov \and N. Yu. Reshetikhin,\jpa 19, 565, 1986

\refis{KiriR87} A. N. Kirillov \and N. Yu. Reshetikhin,\jpa 20, 1565, 1987

\refis{KlassM90} T. R. Klassen \and E. Melzer, \npb 338, 485, 1990

\refis{KlassM91} T. R. Klassen \and E. Melzer, \npb 350, 635, 1991

\refis{KlumBiq} A. Kl\"{u}mper, \eurolett 9, 815, 1989; \jpa 23, 809, 1990

\refis{KlumB90} A. Kl\"{u}mper \and \mtb,\jpa 23, L189, 1990

\refis{KlumBP91} A. Kl\"{u}mper, \mtb \ \and \pap, \jpa 24, 3111--3133, 1991

\refis{KlumP91} A. Kl\"{u}mper \and \pap, \jsp 64, 13--76, 1991

\refis{Klum92c} A. Kl\"{u}mper, unver"offentlichte Rechnungen, (1992)

\refis{KlumZ88} A. Kl\"{u}mper \and J. Zittartz,\zpb 71, 495, 1988

\refis{KlumZ88App} A. Kl\"{u}mper \and J. Zittartz,\zpb 71, 495, 1988,
Appendix A

\refis{KlumZ89} A. Kl\"{u}mper \and J. Zittartz,\zpb 75, 371, 1989

\refis{KlumZ8VM} A. Kl\"{u}mper \and J. Zittartz,\zpb 71, 495, 1988;
\zpb 75, 371, 1989

\refis{KlumSZ89} A. Kl\"{u}mper, A. Schadschneider \and J. Zittartz,
\zpb 76, 247, 1989

\refis{KlumSZMPG} A. Kl\"{u}mper, A. Schadschneider \and J. Zittartz,
\jpa 24, L955-L959, 1991; \zpb 87, 281-287, 1992

\refis{Klum89} \ak, \eurolett 9, 815, 1989

\refis{KlumP92} \ak\  \and \pap, \physica 183A, 304-350, 1992

\refis{Klum92} \ak , \Annp 1, 540, 1992

\refis{Klum93} \ak , \zpb 91, 507, 1993

\refis{Klum92b} \ak, in preparation

\refis{KlumWZ93} \ak, T. Wehner \and J. Zittartz, \jpa 26, 2815, 1993

\refis{Klum94} \ak , in Vorbereitung

\refis{KlumWeh94} \ak\ \and T. Wehner, in Vorbereitung

\refis{Knabe88} S. Knabe, \jsp 52, 627, 1988

\refis{Koma} T. Koma, \ptp 78, 1213, 1987; \bf 81, \rm 783, (1989)

\refis{KorepinS90} V. E. Korepin \and N. A. Slavnov, \npb 340, 759, 1990

\refis{ItsIK92} A. R. Its, A. G. Izergin \and V. E. Korepin, \physica D 54,
351, 1992

\refis{ItsIKS93} A. R. Its, A. G. Izergin, V. E. Korepin \and N. A. Slavnov,
\prl 70, 1704, 1993

\refis{IKR89} A. G. Izergin, V. E. Korepin \and N. Yu. Reshetikhin, \jpa 22,
2615, 1989

\refis{KuniY88} A. Kuniba \and T. Yajima,\jsp 52, 829, 1988

\refis{KuliRS81} P. P. Kulish, N. Yu. Reshetikhin \and E. K. Sklyanin, \lmp
5, 393, 1981

\refis{Kuniba92} A. Kuniba, ``Thermodynamics of the $U_q(X_r^{(1)})$ Bethe
Ansatz System with $q$ a Root of Unity", ANU preprint (1991)

\refis{LeeS88} K. Lee \and P. Schlottmann, \jpcoll 49 C8, 709, 1988

\refis{Lewi58} L. Lewin, Dilogarithms and Associated Functions, MacDonald,
London, 1958

\refis{LiebWu68} E. H. Lieb \and F. Y. Wu, \prl 20, 1445, 1968

\refis{LiebWu72} E. H. Lieb \and F. Y. Wu,
``Phase Transitions and Critical Phenomena,
Vol.1",
\eds C. Domb \and M. S. Green, Academic Press, London 1988

\refis{LutherP74} A. Luther \and I. Peschel, \prb 9, 2911, 1974

\refis{Martins91} M. J. Martins, \prl 22, 419, 1991 and private communication
(1991)

\refis{Muell} E. M\"uller-Hartmann, unpublished results, (1989)

\refis{Muell89} E. M\"uller-Hartmann, unver"offentlichte Ergebnisse, (1989)

\refis{Mura89} J. Murakami, \aspm 19, 399--415, 1989

\refis{Nien87} B. Nienhuis, in Phase Transitions and Critical Phenomena,
Vol.11,
\eds C. Domb \and J.L. Lebowitz, Academic Press, 1987

\refis{NahmRT92} W. Nahm, A. Recknagel \and M. Terhoven, Preprint ``Dilogarithm
identities in conformal field theory'', 1992

\refis{NighB86} M. P. Nightingale \and H. W. J. Bl"ote, \prb 33, 659, 1986

\refis{OwczB87} A. L. Owczarek \and \rjb,\jsp 49, 1093, 1987

\refis{ParkBiq} J. B. Parkinson,\jpc 20, L1029, 1987; \jpc 21, 3793, 1988;
\jphc 8, 1413, 1988

\refis{ParkB85} J. B. Parkinson \and J. C. Bonner, \prb 32, 4703, 1985

\refis{PaczP90} I. D. Paczek \and J. B. Parkinson,\jpcon 2, 5373, 1990

\refis{Pasq87a} V. Pasquier,\npb {285 [FS19]}, 162, 1987

\refis{Pasq87b} V. Pasquier,\jpa 20, {L217, L221}, 1987

\refis{Pasq87c} V. Pasquier,\jpa 20, {L1229, 5707}, 1987

\refis{Pasq88} V. Pasquier,\npb {B295 [FS21]}, 491--510, 1988

\refis{Pear85} \pap,\jpa 18, 3217--26, 1985

\refis{Pear87prl} \pap,\prl 58, 1502--4, 1987

\refis{Pear87jpa} \pap,\jpa 20, 6463--9, 1987

\refis{Pear90ijmpb} \pap,\ijmpb 4, 715--34, 1990

\refis{PearB90} \pap\space \and \mtb, \jsp 60, 77--135, 1990

\refis{PearK87} \pap \and \dk, \jpa, 20, 6471-85, 1987

\refis{PearS88} \pap\space \and K. A. Seaton,\prl 60, 1347, 1988

\refis{PearS89} \pap\space \and K. A. Seaton,\annp 193, 326, 1989

\refis{PearS90} \pap\space \and K. A. Seaton,\jpa 23, 1191--1206, 1990

\refis{Pear91} \pap, Row Transfer Matrix Functional Equations for
$A$--$D$--$E$ Lattice Models,
 to be published, 1991

\refis{PearK91} \pap\space \and A. Kl\"umper, \prl 66, 974, 1991

\refis{Pear92} \pap, \ijmpa 7, Suppl.1B, 791, 1992

\refis{PerkS81} J. H. H. Perk \and C. L. Schultz, \pla 84, 407, 1981

\refis{PensK86} K. Penson \and M. Kolb, \prb 33, 1663, 1986; M. Kolb \and K.
Penson, \jsp 44, 129, 1986

\refis{Resh83jetp} \nyr,\jetp 57, 691, 1983

\refis{Resh83lmp} \nyr, \lmp 7, 205--13, 1983

\refis{Sale88} H. Saleur, Lattice Models and Conformal Field Theories, in
Carg\`ese
School on Common Trends in Condensed Matter and Particle Physics, 1988

\refis{SaleB89} H. Saleur \and M. Bauer, \npb 320, 591--624, 1989

\refis{SaleD87} H. Saleur \and P. di Francesco, Two Dimensional Critical
Models on a
Torus, in Brasov Summer School on Conformal Invariance and String Theory,
1987

\refis{Samuel73} E. J. Samuelson, \prl 31, 936, 1973

\refis{Schlott87} P. Schlottmann, \prb 36, 5177, 1987

\refis{Schlott92} P. Schlottmann, \jpc 4, 7565, 1992

\refis{Schul83} C. L. Schultz, \physica 122A, 71, 1983

\refis{SeatP89} K. A. Seaton \and \pap\space, \jpa 22, 2567--76, 1989

\refis{Strog79} Yu. G. Stroganov, \pla 74, 116, 1979

\refis{Suth70} B. Sutherland, \jmp 11, 3183, 1970

\refis{Suth75} B. Sutherland, \prb 12, 3795, 1975

\refis{Suth} B. Sutherland, in Exactly Solved Problems in Condensed Matter and
Relativistic Field Theory (Lecture Notes in Physics 242), \eds
B. S. Shastry, S. S. Iha \and V. Sigh, Springer, Berlin, 1982

\refis{Suzuki85} M. Suzuki, \prb 31, 2957, 1985

\refis{SuzukiI87} M. Suzuki \and M. Inoue, \ptp 78, 787, 1987

\refis{Suzuki87} M. Suzuki, in ``Quantum Monte Carlo Methods in
Equilibrium and Nonequilibrium Systems",
\edi M. Suzuki, Springer Verlag, 1987

\refis{SuzukiAW90} J. Suzuki, Y. Akutsu \and M. Wadati, \jpj 59, 2667-2680,
1990

\refis{SuzukiNW92} J. Suzuki, T. Nagao \and M. Wadati, \ijmpb 6, 1119, 1992

\refis{Tak71} M. Takahashi, \ptp 46, 401, 1971

\refis{TakTBA} M. Takahashi, \ptp 46, 401, 1971; \ptp 50, 1519, 1973

\refis{Tak91} M. Takahashi, \prb 43, 5788, 1991; \prb 44, 12382, 1991

\refis{Tak91a} M. Takahashi, \prb 43, 5788, 1991

\refis{Tak91b} M. Takahashi, \prb 44, 12382, 1991

\refis{TempL71} H. N. V. Temperley \and E. H. Lieb, \prs 322, 251, 1971

\refis{Tetel82} M. G. Tetel'man, \jetp 55, 306, 1982

\refis{TruS83} T. T. Truong \and K. D. Schotte, \npb 220, 77, 1983

\refis{Tsun91} H. Tsunetsugu, \jpj 60, 1460, 1991

\refis{vonGR87} G. von Gehlen \and V. Rittenberg, \jpa 20, 227, 1987

\refis{WadaDA89} M. Wadati, T. Deguchi \and Y. Akutsu, \prep 180, 247--332,
1989

\refis{Woyn87} F. Woynarovich, \prl 59, 259, 1987

\refis{WoynE87} F. Woynarovich \and H.-P. Eckle, \jpa 20, L97, 1987

\refis{Yang69} C. N. Yang \and C. P. Yang, \jmp 10, 1115, 1969

\refis{Yang66} C. N. Yang \and C. P. Yang, \pr 147, 303, 1966; 150, 321

\refis{Yang62} C. N. Yang, \rmp 34, 691, 1962

\refis{Yang67} C. N. Yang, \prl 19, 1312, 1967

\refis{CPYang67} C. P. Yang, \prl 19, 586, 1967

\refis{YangG89} C. N. Yang \and M. L. Ge (Editors), Braid Group, Knot Theory
and Statistical
Mechanics, World Scientific, Singapore, 1989

\refis{ZamoF80} A. B. Zamolodchikov \and V. Fateev, \sjnp 32, 198, 1980

\refis{Zamo80} A. B. Zamolodchikov, \jetp 52, 325, 1980; \cmp 79, 489, 1981

\refis{Zamo91} Al. B. Zamolodchikov, \plb 253, 391--4, 1991; \npb 358,
497--523, 1991

\refis{Zamo91a} Al. B. Zamolodchikov, \plb 253, 391--4, 1991

\refis{Zamo91b} Al. B. Zamolodchikov, \npb 358, 497--523, 1991

\refis{ZhangR88} F. C. Zhang and T. M. Rice, \prb 37, 3759, 1988

\endreferences

\vskip0.5cm

\section*{Figure captions}

\begin{itemize}

\item[Figure 1:] a) Depiction of the density dependence of the groundstate
energy per site $E_0$ and the exponent $\beta$ of the pair correlation
function for values $\eps=+1$, $\gamma=0.5$, and $\eta=$ 0.6, 1, 2, 4.
b) Similar to a) for $\eps=+1$, $\eta=1$, and $\gamma=$ 0.1, 0.3, 0.5, 0.7,
0.9.

\item[Figure 2:] Similar to 1a) and b) but for $\eps=-1$.
\end{itemize}

\newpage

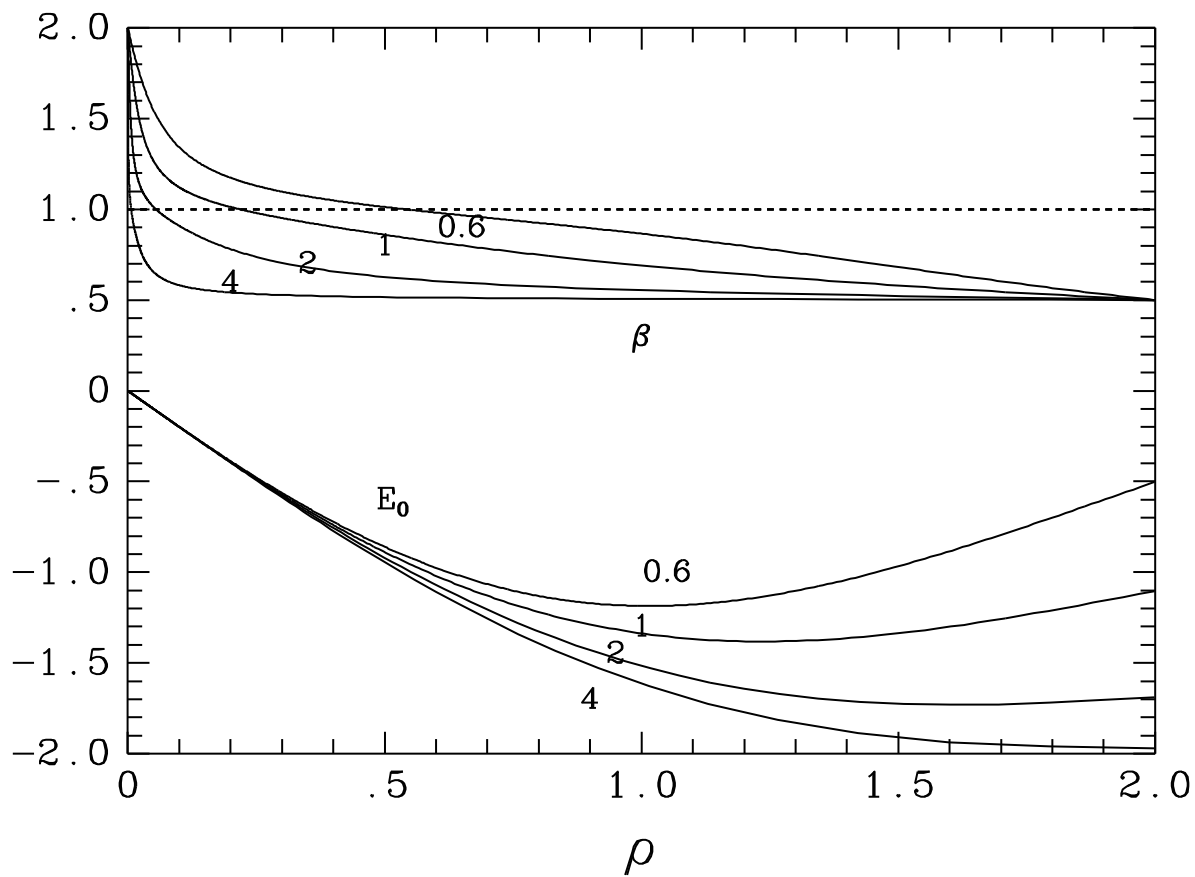
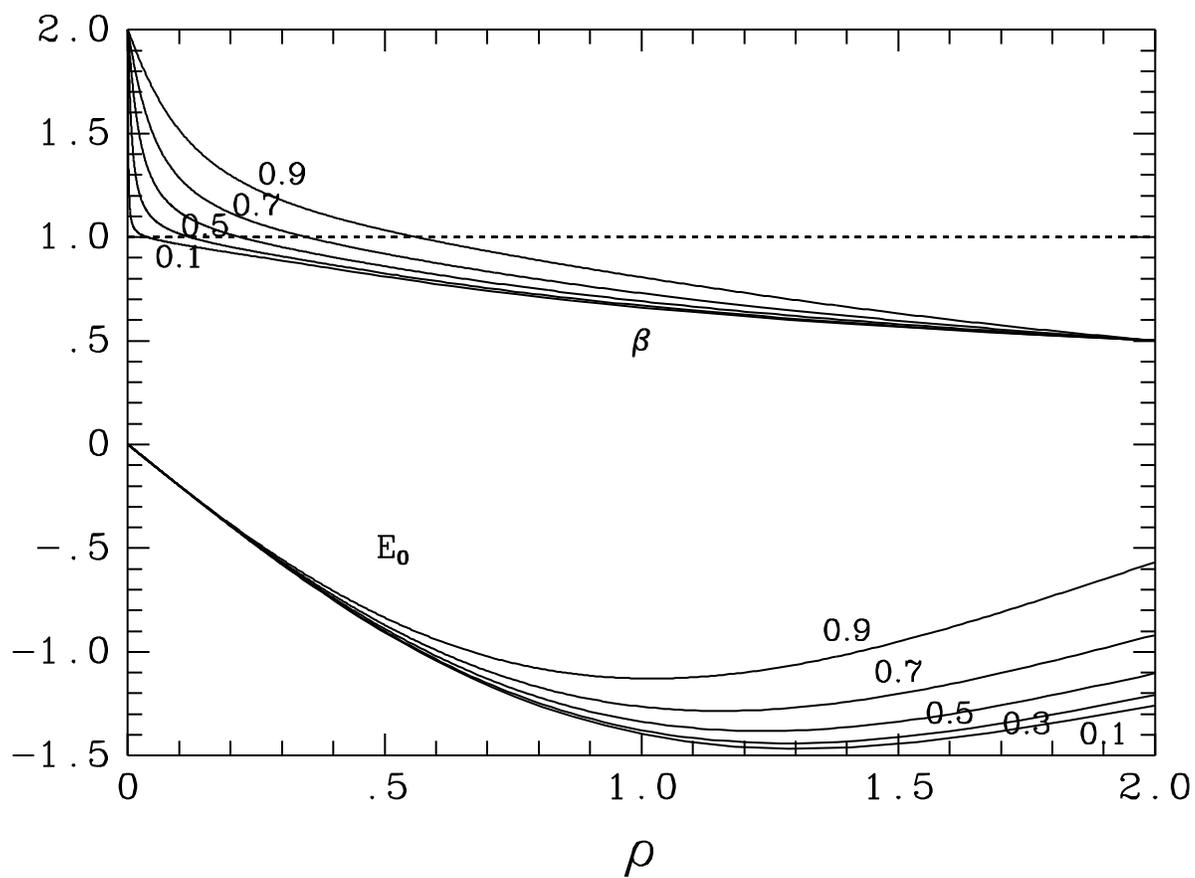
\end{document}